\documentclass[runningheads]{llncs}
\usepackage[T1]{fontenc}
\usepackage{graphicx}
\usepackage{url} 
\usepackage[utf8]{inputenc} 
\usepackage{booktabs} 
\usepackage{hyperref} 
\usepackage{xcolor}
\begin{document}
\title{Measuring Modern Phishing Tactics: A Quantitative Study of Body Obfuscation Prevalence, Co-occurrence, and Filter Impact}
%
%

\author{Antony Dalmiere\inst{1}\orcidID{0009-0009-0019-112X} \and Zheng Zhou\inst{1}\orcidID{0009-0006-0545-5639} \and  
Guillaume Auriol\inst{1,2}\orcidID{0009-0001-2775-5345} \and
Vincent Nicomette\inst{1,2}\orcidID{0000-0001-9482-004X} \and Pascal Marchand\inst{3}\orcidID{0000-0002-4190-8213}}
\authorrunning{Dalmiere et al.}
%
\institute{CNRS, LAAS, 7 avenue du colonel Roche, F-31400,\\ \email{[firstname.lastname]@laas.fr}\and
 Université de Toulouse, INSA, F-31400\\
\and
LERASS, Université of Toulouse, France\\
\email{pascal.marchand@iut-tlse3.fr}}
\maketitle              
\begin{abstract}
Phishing attacks frequently use email body obfuscation to bypass detection filters, but quantitative insights into how techniques are combined and their impact on filter scores remain limited. This paper addresses this gap by empirically investigating the prevalence, co-occurrence patterns, and spam score associations of body obfuscation techniques. Analyzing 386 verified phishing emails, we quantified ten techniques, identified significant pairwise co-occurrences revealing strategic layering like the presence of text in image with multipart abusing, and assessed associations with antispam score using multilinear regression. Text in Image (47.0\%), Base64 Encoding (31.2\%), and Invalid HTML (28.8\%) were highly prevalent. Regression (R²=0.486, p<0.001) linked Base64 Encoding and Text in Image with significantly antispam evasion (p<0.05) in this configuration, suggesting potential bypass capabilities, while Invalid HTML correlated with higher scores. These findings establish a quantitative baseline for complex evasion strategies, underscoring the need for multi-modal defenses against combined obfuscation tactics.

\end{abstract}

\section{Introduction}
\label{sec:introduction}
Phishing remains one of the most pervasive and damaging cyber threats confronting individuals and organizations globally. As a primary vector for credential harvesting, malware dissemination, financial fraud, and espionage \cite{ibm2024,verizon2024}, phishing campaigns inflict substantial economic and operational costs annually \cite{getastra2025,cisa2023}. The enduring success of these attacks is intrinsically linked to the attackers' adeptness at circumventing sophisticated, multi-layered security defenses, including anti-spam filters, URL reputation services, sandboxing environments, and advanced content analysis engines. Central to this evasion strategy is the deployment of obfuscation techniques within the email body. These techniques are meticulously crafted to manipulate email content and structure, rendering malicious elements less detectable by automated systems while preserving — or even enhancing — their deceptive appearance and functionality for the intended human recipient.

Modern phishing campaigns frequently employ a combination of obfuscation techniques, strategically layered to bypass diverse detection heuristics and maximize the probability of reaching the target's inbox and deceiving automated filters. However, to the best of our knowledge, a comprehensive empirical measurement quantifying how these techniques are combined and their potential association with filter evasion is lacking in the existing literature.

This paper aims to bridge this critical gap by presenting a data-driven quantitative investigation into obfuscation techniques within real-world phishing emails. Our central research questions are:

\begin{itemize}
\item How frequently are different obfuscation techniques used in phishing emails? 
\item How are these techniques strategically combined? 
\item What is the statistical association between the presence of these techniques and the spam scores assigned by a baseline filter?
\end{itemize}
We hypothesize that analyzing the prevalence, statistical co-occurrence patterns, and filter score associations provides a more nuanced understanding of attacker evasion strategies.

To test this hypothesis, we conduct a detailed analysis of a corpus comprising 386 manually verified phishing emails. Our quantitative methodology moves beyond simple descriptive statistics by employing:
\begin{itemize}
    \item \textbf{Prevalence Analysis:} Establishing baseline frequencies for individual body obfuscation techniques.
    \item \textbf{Co-occurrence Analysis:} Identifying statistically significant co-occurrences between pairs of techniques to quantify which techniques are deliberately used in concert.
    \item \textbf{Effectiveness Association Analysis:} Utilizing multilinear regression to assess the statistical association between the presence of each technique and the spam score assigned by SpamAssassin, offering a preliminary indicator of potential filter bypass effectiveness.
\end{itemize}

To summarize, our contribution is threefold: 

\begin{enumerate}
\item  We provide robust quantitative evidence of the prevalence and common combinations of body obfuscation techniques in phishing emails, establishing an empirical baseline.
\item We quantify statistically significant co-occurrence patterns, revealing frequently employed strategic pairings.
\item We offer an initial quantitative assessment of how individual techniques correlate with spam scores from a widely used filter (SpamAssassin), identifying techniques associated with score modification in our specific test configuration.
\end{enumerate}

The remainder of this paper is organized as follows. Section~\ref{sec:related_work} reviews prior work on email obfuscation and detection. Section~\ref{sec:methodology} details our data collection process, the methodology for identifying body obfuscation techniques, and the quantitative analysis methods employed. Section~\ref{sec:results} presents the core quantitative findings, including technique prevalence, co-occurrence analysis, and the effectiveness association results. Section~\ref{sec:discussion} interprets these results, discusses their implications for understanding attacker behavior and improving defenses, and acknowledges limitations. Finally, Section~\ref{sec:conclusion} summarizes our key contributions and outlines directions for future research.

\section{Related Work}
\label{sec:related_work}

Early work \cite{dhiman2016obfuscation}, demonstrated how specific techniques like homograph substitutions could bypass digest and signature-based detection systems in major webmail providers, highlighting the potential impact of even single, focused methods.

Other research analyzes phishing detection failures at semantic  level. For instance, Champa et al. \cite{champa2024} performed an in-depth qualitative analysis of emails misclassified by machine learning models, identifying semantic reasons for failure, such as mimicking legitimate communication styles or exploiting subject line characteristics. While valuable, this type of analysis often occurs after initial email parsing and feature extraction, potentially overlooking evasion tactics designed to disrupt these earlier stages. Our work complements this by focusing specifically on obfuscation techniques that challenge the email processing pipeline at or before the parsing and feature extraction phase, aiming to make malicious content inaccessible or misleading to subsequent semantic analyses.

Furthermore, as noted by \cite{salloum2022a} , existing phishing research \cite{shahriar2022aa} frequently relies on established datasets such as PhishTank \cite{ubcPhishTank2017}, the SpamAssassin public corpus \cite{spamassassinCorpus}, the ENRON dataset \cite{enronEmailDataset}, TREC datasets \cite{trec2005spam,cormack2006trecspam,trecspamcorpus2007}, Phishload \cite{phishload_maurer}, and the Nazario Corpus \cite{NazarioPhishingCorpus}. However, a common practice during the curation of these datasets, or in studies utilizing them, involves excluding emails deemed corrupted or anomalous. This often leads to the removal of messages exhibiting significant formatting anomalies, non-standard structures which indicate heavy obfuscation. While such curation facilitates model training on cleaner data, it can consequently lead to detection models that perform poorly when encountering heavily obfuscated emails in real-world scenarios, as these messages represent out-of-distribution samples. Our work directly addresses this gap by specifically analyzing the techniques employed to create such challenging, obfuscated emails.

Therefore, our research addresses a critical gap by providing empirical data on the usage patterns and co-occurrence of body obfuscation techniques. This quantitative approach offers novel insights into attacker strategies designed specifically to evade automated detection mechanisms.

\section{Methodology}
\label{sec:methodology}

Our research methodology was designed to enable a rigorous quantitative examination of obfuscation patterns within phishing emails. The process involved data acquisition, manual verification, and systematic feature extraction.

\subsection{Data Collection and Preparation}

Given the enormous volume of phishing emails reported during the collection period (2024), estimated to easily exceed the million \cite{keepnetlabs2025phishing,APWG2024Q3,baker2025phishing}, 
we adopted a statistically grounded sampling strategy to ensure meaningful and generalizable findings while maintaining practical feasibility.
We aimed to estimate the prevalence of obfuscation techniques with a margin of error no greater than $\pm$5\% at a 95\% confidence level. Assuming maximum uncertainty (i.e., a technique may be present in approximately 50\% of emails), the required sample size $n$ can be calculated using the Cochran’s sample size formula \cite{nanjundeswaraswamy2021}:

\begin{equation}
\label{eq::standard}
n = \frac{Z^2 \cdot p \cdot (1 - p)}{E^2},
\end{equation}

where $Z$ is the z-score for the desired confidence level ($1.96$ for 95\%), $p$ is the estimated proportion ($0.5$ for maximum variance), and $E$ is the acceptable margin of error ($0.05$). Plugging in the values, we obtain:

\begin{equation}
\label{eq::standard_value}
n = \frac{(1.96)^2 \cdot 0.5 \cdot 0.5}{(0.05)^2} \approx 384.
\end{equation}

To ensure robustness across our co-occurrence and regression analyses, we slightly exceeded this threshold. Our final dataset contains 386 manually verified phishing emails, meeting the statistical requirements while enabling detailed analysis of obfuscation strategies.
These 386 emails were collected during 2024 via SignalSpam\footnote[1]{https://www.signal-spam.fr/}, a French organization that facilitates user reporting of phishing, spam, and scams. Emails received through this platform are initially flagged by users, and each message in our final corpus underwent manual verification by our team to confirm its phishing nature. All data handling adhered to strict ethical guidelines, including the anonymization of potential Personally Identifiable Information (PII) and secure data storage.

\subsection{Identification of Body Obfuscation Techniques}
Each of the 386 verified phishing emails was meticulously inspected by a trained analyst. Then a subset was checked by an other specialist. The inspection involved examining both the raw source code (typically within the `.eml` file) and the rendered output as displayed by common email clients. Due to the limited prior work systematically cataloging combined obfuscation techniques in this manner, we developed a classification scheme based on observed patterns and established evasion principles. The identification process relied on identifying discrepancies and anomalies based on the following primary criteria:
\begin{enumerate}
    \item \textbf{Source vs. Rendered Discrepancy:} Significant differences between the underlying source code (e.g., HTML, MIME structure) and the visually rendered content presented to the user.
    \item \textbf{Suspicious Character Usage:} Presence of non-standard characters, invisible characters, or character encodings (like Base64) applied in unusual contexts (e.g., encoding short strings, URLs).
    \item \textbf{Structural Anomalies:} Deviations from standard email formatting specifications (e.g., RFC 5322 \cite{resnick2008}, MIME RFC 2045/2046 \cite{freed1996}) or web standards (e.g., W3C HTML specifications), such as malformed HTML or misuse of multipart structures.
\end{enumerate}
Instances where the classification of a technique was ambiguous were reviewed and adjudicated by a second expert analyst to ensure consistency and accuracy in labeling. This process resulted in the identification and categorization of the techniques described in Section~\ref{sec:results}.

\section{Results}
\label{sec:results}
This section presents the quantitative findings derived from the manual analysis of the 386 phishing emails. We first detail the prevalence of individual body obfuscation techniques identified in the corpus, providing a baseline understanding before examining their combinations and associations with SpamAssassin scores.

\subsection{Exploratory Analysis of Obfuscation Techniques}
\begin{figure}[htbp]
    \centering
    \includegraphics[width=\textwidth]{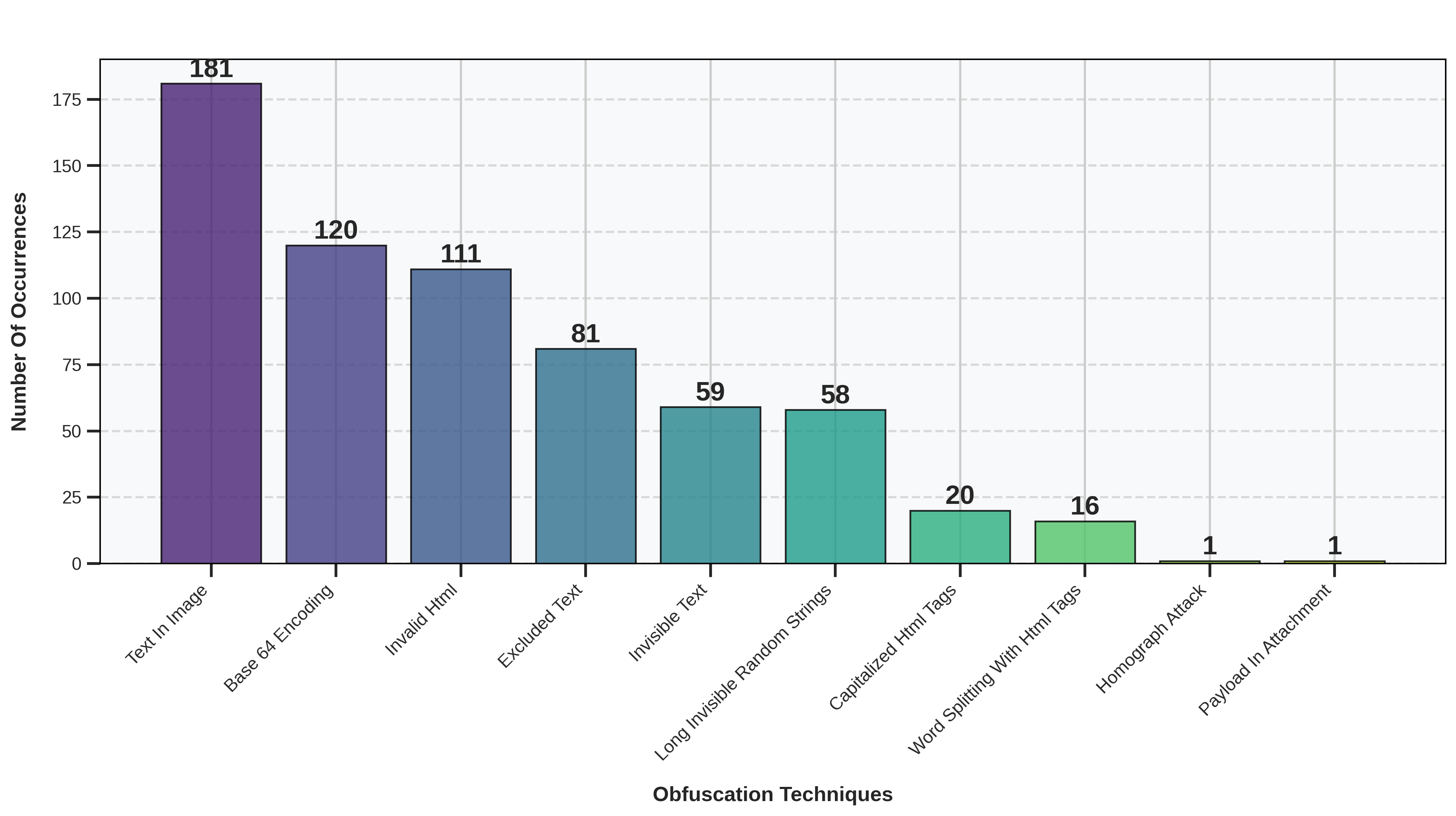}
    \caption{Prevalence of identified body obfuscation techniques in the corpus (N=386).}
    \label{fig:usagebar}
\end{figure}

The following subsections elaborate on each identified technique, ordered by frequency of occurrence. Usage frequencies are summarized in Figure~\ref{fig:usagebar}.

\subsubsection{Text in Image}
The most frequently observed obfuscation strategy, identified in 181 emails (47.01\% of the corpus), involves rendering the primary message content as an image. In these instances, the core textual information — including persuasive language, branding elements, calls to action, and even visually depicted hyperlinks — is embedded directly within an image file (e.g., JPEG, PNG, GIF). Consequently, the email body, when analyzed as raw text or standard HTML, may contain minimal machine-readable content, often limited to the \texttt{<img>} tag referencing the image file. This method fundamentally circumvents conventional text-based analysis pipelines, including keyword filtering, regular expression matching, and Natural Language Processing (NLP) techniques commonly employed by anti-spam and anti-phishing systems. Effective defense against this technique necessitates the integration of Optical Character Recognition (OCR) capabilities. However, OCR presents its own challenges: accuracy can be degraded by image quality, complex backgrounds, non-standard fonts, or deliberate manipulations like text distortion. This last case was not observed. Furthermore, deploying OCR at scale imposes significant computational overhead. Attackers may also employ hybrid approaches, mixing small amounts of real text with image-based content to further complicate creation of heuristics detecting this technique.

\subsubsection{Base64 Encoding}
Base64 encoding was detected in 120 emails (31.17\% of the analyzed set). This technique represents text strings (such as URLs, JavaScript snippets, or HTML segments) using only printable ASCII characters. While Base64 encoding is a standard mechanism with legitimate uses in email (e.g., embedding inline images, handling non-ASCII characters in headers), its application in these phishing emails appears strategically aimed at evasion. The primary hypothesis is that encoding malicious components serves to bypass rudimentary content filters or signature-based detection engines that may not automatically decode and inspect Base64-encoded segments, potentially due to performance constraints or assumptions about encoded content. Although decoding Base64 is computationally trivial for sophisticated security systems, its presence necessitates an additional processing step. Attackers might also employ multiple layers of encoding or combine Base64 with other obfuscation methods even-tough it's scare (detailed in \ref{subsec:cooccurrence}). Distinguishing malicious use often relies on context, such as encoding unusually short strings associated with malicious activity rather than legitimate large binary objects.

\subsubsection{Invalid HTML}
A substantial fraction of the emails, 111 instances (28.83\%), exhibited deliberately malformed HTML structures. This category encompasses huge and deliberate deviations from W3C standards, including improperly nested tags (e.g., \texttt{<b><i>text</b></i>}), unclosed tags, closing tags without corresponding opening tags, or redundant closing tags. The likely intent is multi-faceted: primarily, it aims to disrupt, confuse, or potentially crash automated HTML parsers used by security tools during content extraction. If a parser fails or generates an incorrect Document Object Model (DOM), subsequent analysis of content, links, or scripts may be incomplete or inaccurate. Attackers might exploit specific vulnerabilities or error-handling quirks in different parsing engines. This technique presents a trade-off: while impeding automated analysis, excessively broken HTML might also cause rendering issues in the victim's email client as observed in dataset. Modern email clients are often highly tolerant of HTML errors, potentially rendering malicious content correctly for the user while still hindering stricter security parsers.

\subsubsection{Excluded Text (Multipart Abuse)}
Present in 81 emails (21.04\%), this technique leverages the \texttt{multipart/alternative} MIME structure evasively. The MIME standard allows emails to include multiple versions of content (typically plain text and HTML), enabling the client to choose the best format. Attackers exploit this by placing malicious content (deceptive narratives, phishing links) exclusively within the HTML part, while populating the corresponding plain text part with different, often innocuous or unrelated content (e.g., boilerplate text). The evasion strategy relies on the possibility that some email clients or, more critically, intermediate security filters might preferentially analyze or fallback to the simpler \texttt{text/plain} version due to security policies, performance optimizations, or robustness measures against malformed HTML. By examining only the benign plain text part, such systems would miss the malicious payload delivered via the HTML alternative, highlighting the importance for security systems to consistently analyze the content intended for final rendering.

\subsubsection{Invisible Text (Meaningful Content)}
Detected in 59 emails (15.32\%), this method involves hiding substantial amounts of legitimate-seeming text within the email's HTML structure, rendering it invisible to the recipient using techniques like CSS (\texttt{display:none}, \texttt{visibility:hidden}), zero font size, matching font color to the background or in non-displayable tag. Crucially, the hidden content is typically meaningful text, often copied from legitimate sources. In the dataset, transactional emails, news articles and Wikipedia passages were observed. The hypothesized goal is primarily an adversarial attack targeting sophisticated detection models, particularly those relying on semantic analysis or machine learning classifiers trained on text features (e.g., TF-IDF, Neural networks). By injecting large volumes of benign, coherent text, attackers likely aim to manipulate the email's overall feature representation, skewing it towards a legitimate classification and effectively drowning out smaller malicious signals. This challenges models trained to identify phishing based on semantic.

\subsubsection{Long Invisible Random Strings}
Observed in 58 emails (15.06\%), this technique involves inserting lengthy strings of apparently random characters, rendered invisible using methods similar to those for meaningful invisible text. These strings differ from Base64 encoding (lack of standard structure) and meaningful invisible text (lack of semantic coherence). We hypothesize these random strings primarily serve as adversarial attacks targeting statistical analysis and some machine learning techniques. Their presence can perturb statistical properties relied upon by detection models, such as character frequency distributions, n-gram analyses, or document entropy calculations. Furthermore, because these strings often lack spaces, they might be treated as single, extremely long tokens, potentially disrupting tokenization processes, vocabulary assumptions, feature extraction mechanisms in ML models or even parsing. In some cases, excessively long strings might also aim to trigger buffer overflows or cause excessive resource consumption in less robust parsing components, acting as a denial-of-service vector against the detection system.

\subsubsection{Capitalized HTML Tags}
This relatively straightforward technique was observed in 20 emails (5.19\%). It involves deviating from the standard practice of using lowercase letters for HTML element names, instead employing uppercase (e.g., \texttt{<DIV>}) or mixed-case letters (e.g., \texttt{<sPaN>}). This manipulation serves no legitimate purpose. The hypothesized intent is to evade even less sophisticated parsers or signature-based detection rules relying on exact, case-sensitive string matching for HTML tags. While modern browsers and email clients are generally case-insensitive regarding HTML tags, security tools or intermediate gateways might implement stricter parsing rules or use case-sensitive regular expressions, which could be bypassed by this capitalization change. Its relatively low prevalence suggests it may target older or less robust filtering systems.

\subsubsection{Word Splitting with HTML Tags}
Found in 16 emails (4.16\%), this technique disrupts the contiguity of specific words by inserting HTML tags directly within them. Formatting tags with no visual effect are used (e.g., \texttt{Bit<span></span>coin}, \texttt{P<font>a</font>ssword}). This splitting serves no discernible visual purpose. The primary hypothesized goal is to circumvent keyword-based detection systems and content filters scanning for blacklisted terms. By breaking the character sequence forming the target keyword, the technique prevents a direct string match. This might also challenge certain NLP techniques relying on standard tokenization, as the word is fragmented in the HTML source. Effective detection requires parsers capable of reconstructing textual content by appropriately handling these interspersed tags.

\subsubsection{Homograph Attack}
Observed with very low frequency (1 email, 0.26\%) in this dataset, the homograph attack leverages Unicode characters that are visually identical or extremely similar (homographs) to standard ASCII characters. For instance, the Latin 'a' might be replaced with the Cyrillic 'a'. While appearing identical to a human, these characters have distinct underlying Unicode code points. Consequently, automated systems performing string comparisons based on code points perceive the altered word (e.g., in a domain name) as different from the original ASCII version. This technique is also used in URL filtering, brand protection, and credential harvesting by enabling visually deceptive domain names or disguised links \cite{dhiman2016obfuscation}. Its rarity here might indicate selective use or lower prevalence during the collection period, but its potential for high-impact deception remains significant.

\subsubsection{Payload in Attachment}
Also exhibiting very low prevalence (1 email, 0.26\%), this strategy minimizes malicious indicators within the email body, placing the core payload in an attached file. The email body might appear benign or empty. The attachment contains the actual phishing lure : an another embedded email. This shifts the detection burden to attachment analysis systems. Attackers exploit the fact that attachment scanning can be resource-intensive and subject to limitations (file size, timeouts, unsupported types, etc.). A clean-looking email body might receive a lower initial risk score, potentially passing preliminary checks before attachment analysis is completed. The low frequency suggests this was not a preferred method in this email set, possibly due to user suspicion towards unexpected attachments or improved attachment scanning.

\subsection{Technique Co-occurrence Analysis}
\label{subsec:cooccurrence}
\begin{figure}[htbp]
    \centering
    \includegraphics[width=\textwidth]{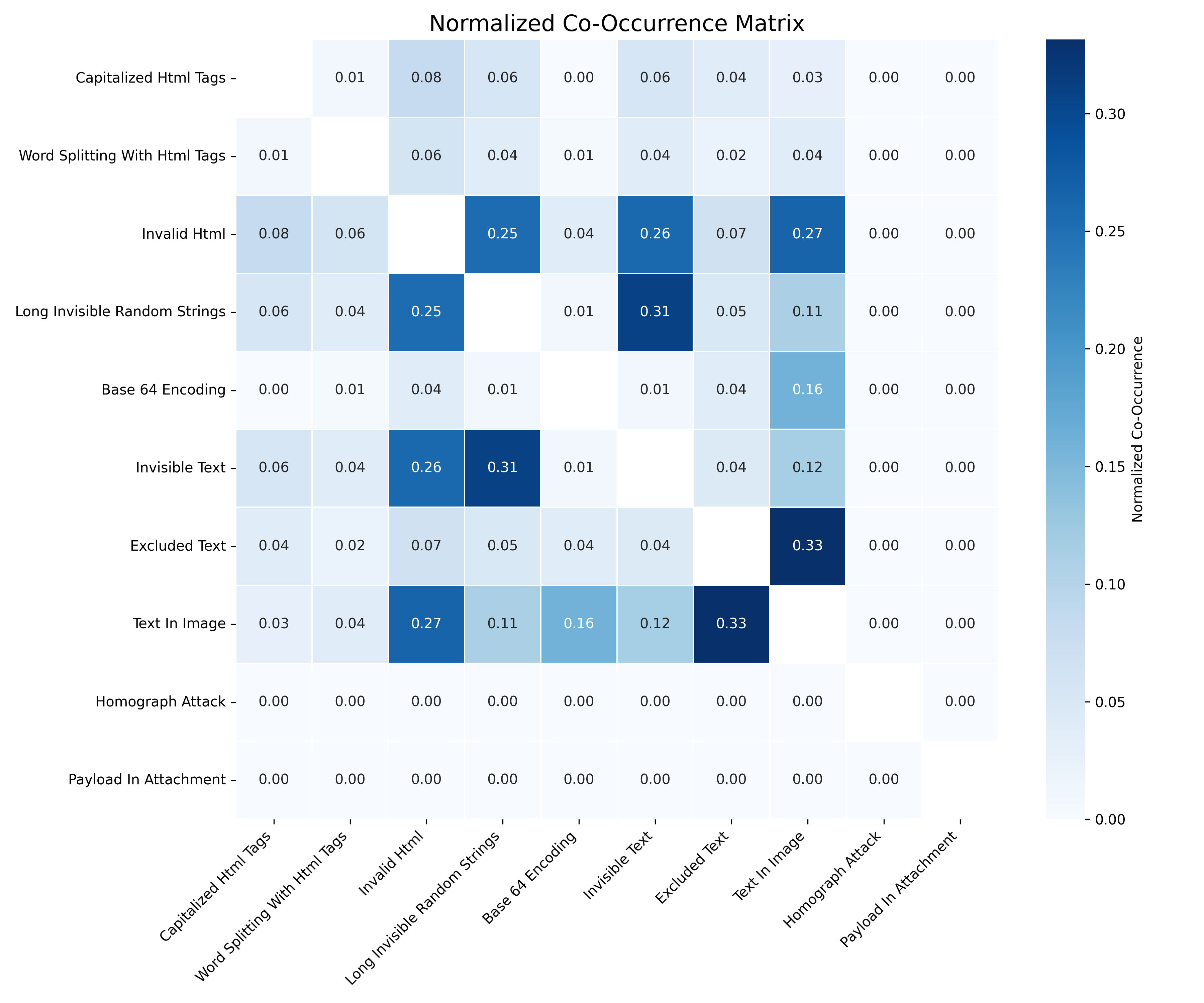}
    \caption{Normalized co-occurrence matrix of obfuscation techniques}
    \label{fig:cooccurrence_matrix}
\end{figure}

Beyond individual prevalence, understanding how attackers strategically combine obfuscation techniques is crucial. We analyzed pairwise co-occurrence patterns, visualized in the normalized co-occurrence matrix (Figure~\ref{fig:cooccurrence_matrix}). This matrix depicts the proportion of emails containing technique $i$ that also contain technique $j$. Self-occurrences (diagonal) are excluded.

The analysis of co-occurrence values provides insight into attacker strategies. Given the ten distinct obfuscation techniques identified highly present co-occurrences between specific pairs may indicate non-random, deliberate combinations employed by attackers. These patterns reveal frequently used strategic pairings, suggesting common evasion playbooks.

Several strong co-occurrences highlight common attacker strategies. The most notable is between \textbf{Excluded Text (Multipart Abuse)} and \textbf{Text in Image} with a normalized co-occurrence of 0.33. This pairing represents a potent multi-layered evasion strategy: the image bypasses direct text analysis, while multipart abuse attempts to deceive filters analyzing the benign plain text part instead of the malicious HTML part containing the image. This combination targets both content inspection and structural analysis defenses.

Another significant cluster involves techniques hiding or distorting textual content. \textbf{Long Invisible Random Strings} frequently appears with \textbf{Invisible Text (Meaningful Content)} with a normalized co-occurrence of 0.31. This suggests attackers may employ both noise injection (random strings) and semantic camouflage (meaningful hidden text) injecting high amount of text which will only be meaningful for filters.

\textbf{Invalid HTML} emerges as a foundational disruption technique, exhibiting notable co-occurrence with multiple content-hiding methods: \textbf{Text in Image} (0.27), \textbf{Invisible Text} (0.26), and \textbf{Long Invisible Random Strings} (0.25). This pattern strongly suggests attackers often attempt to break or confuse HTML parsers first, potentially degrading subsequent content analysis before deploying image-based or text-hiding obfuscations.

Moderate co-occurrences, such as \textbf{Base64 Encoding} with \textbf{Text in Image}  have a normalized co-occurrence of 0.16, indicate pairings employed with some regularity but less dominant than primary clusters. Here, encoding might be used for image source links or auxiliary scripts alongside the main image payload. Numerous other pairs exhibit very low co-occurrence values (< 0.05), suggesting they are not statistically preferred strategies within this dataset. The near-zero co-occurrence involving the rare \textbf{Homograph Attack} and \textbf{Payload in Attachment} aligns with their minimal individual prevalence.

Collectively, these patterns provide quantitative evidence that phishing obfuscation often involves deliberate, synergistic pairings chosen to counter anticipated defenses across different layers of email analysis.

\subsection{Effectiveness Analysis: Associations with SpamAssassin Scores}
\label{subsec:effectiveness}
To gain preliminary insights into the potential impact of these techniques on automated filters, we analyzed the association between the presence of specific obfuscation techniques and the resulting spam score assigned by SpamAssassin (v4.0.1, rules updated as of 06/04/2025). We emphasize that this analysis targets a specific, widely-used open-source filter configuration at a fixed point in time; results may differ significantly for proprietary enterprise filters or future SpamAssassin updates. Nonetheless, this provides a valuable empirical benchmark. To obtain these scores, SpamAssassin was first deployed within a Docker container to ensure a controlled and reproducible environment. Following deployment, the software and its rule sets were updated to reflect the state as of the specified date (06/04/2025). The spam score for each individual email in the corpus (N=386) was then generated by executing the standard SpamAssassin command-line tool against the raw email file: \texttt{spamassassin <path to mail>}. We employed multilinear regression, a statistical technique used to model the relationship between a single dependent variable and multiple independent variables. The general form of a multilinear regression equation is:
\[ Y = \beta_0 + \beta_1 X_1 + \beta_2 X_2 + \dots + \beta_p X_p + \epsilon \]
where $Y$ is the dependent variable, $X_1, \dots, X_p$ are the independent variables, $\beta_0$ is the intercept, $\beta_1, \dots, \beta_p$ are the regression coefficients representing the change in $Y$ for a one-unit change in the corresponding $X$, and $\epsilon$ is the error term representing unexplained variability. In our analysis, the SpamAssassin score served as the dependent variable ($Y$). The presence (+1) or absence (-1) of each identified obfuscation technique served as independent binary predictor variables ($X_i$). This allowed the model to estimate the association of each technique with the final score while controlling for the presence of other techniques in the email.

Standard linear regression relies on certain statistical assumptions. To ensure our findings regarding coefficient significance are robust without needing to explicitly test all underlying assumptions (like normality of distributions), we employed a bootstrap resampling method described by Efron \cite{efron1994}. This involved generating new datasets (B=50000) by repeatedly sampling emails, with minor random replacements, from our original set of 386. We then fitted the same multilinear regression model to each of these simulated datasets. The coefficients, p-values, and 95\% confidence intervals reported in Table~\ref{tab:regression_results} are derived from the distribution of results across the aggregation of these datasets. This approach provides reliable estimates of statistical significance. The overall model fit metrics (R²) is reported on the original dataset.

The initial  model explained a notable portion of the variance in scores (R² = 0.486, F(10, 375) = 35.43, p < 0.001). The R² value indicates that nearly half of the variability in SpamAssassin scores in our dataset can be associated with the presence or absence of the measured body obfuscation techniques. The low p-value confirms the overall statistical significance of the model, meaning these techniques collectively have a non-random relationship with the assigned scores. However, this also implies that factors beyond the measured body techniques (e.g., header features, link reputation, textual nuances not captured by our binary features) account for the remaining variance.

The bootstrap analysis results (Table~\ref{tab:regression_results}) highlight statistically significant associations (p < 0.05) for several techniques, based on their bootstrap-derived p-values and 95\% confidence intervals:

\begin{itemize}
    \item \textbf{Techniques Associated with Score Reduction:} The presence of {\bf Base64 Encoding} (coefficient $\approx$ -0.97, p < 0.001, 95\% CI [-1.29, -0.67]) and {\bf Text in Image} (coefficient $\approx$ -0.29, p = 0.046, 95\% CI [-0.62, 0.01]) were associated with statistically significant decreases in the SpamAssassin score.  The confidence intervals for both coefficients lie entirely or almost entirely below zero, reinforcing the significance of this negative association.  This suggests these techniques, within this model and dataset, tended to lower the perceived spamminess according to SpamAssassin's rules at the time of analysis, aligning with the hypothesis that encoding content or placing it within images can bypass certain text-based rules.
    \item \textbf{Technique Associated with Score Increase:} Conversely, the use of {\bf Invalid HTML} was associated with a statistically significant increase in the spam score (coefficient $\approx$ 1.11, p < 0.001, 95\% CI [0.65, 1.54]).  The confidence interval for this coefficient is entirely above zero.  This strongly indicates that, rather than aiding evasion against this filter, deviations from standard HTML structure were actively penalized by this version of SpamAssassin, likely triggering specific rules flagging malformed messages.
    \item \textbf{Techniques without Statistically Significant Association:} Several techniques, including {\bf Long Invisible Random Strings}, {\bf Invisible Text}, {\bf Excluded Text}, {\bf Word splitting with HTML Tags}, {\bf Capitalized HTML Tags}, {\bf Homograph attack}, and {\bf Payload in attachment}, did not show a statistically significant relationship with the spam score (p > 0.05) in this multivariable model after controlling for other factors.  For these techniques, the bootstrap-derived 95\% confidence intervals contain zero, indicating that the observed association could plausibly be due to chance.  Their lack of individual significance here does not definitively rule out an effect, particularly in combination or against different filters.
\end{itemize}

These findings must be interpreted cautiously. The linear regression identifies associations, not causal proof of evasion effectiveness. The model assumes linear, additive effects, potentially neglecting complex interactions or non-linear filter responses. The binary encoding also simplifies technique implementation nuances. Therefore, these results serve as quantitative indicators of how these techniques relate to scores from a specific anti-spam engine configuration, providing empirical grounding for understanding potential challenges to rule-based analysis, but warranting further investigation against diverse and evolving filters.

\begin{table}[htbp!]
\centering
\caption{Multilinear Regression Results: Association of Techniques with SpamAssassin Score}
\label{tab:regression_results}
\begin{tabular}{@{}lrrr@{}} 
\toprule 
Technique                       & Coefficient & \textit{p}-value & 95\% CI \\ 
\midrule
(Intercept)                     &  4.23**     & 0.027      & [1.50, 5.29] \\ 
{\bf Base64 Encoding}                 & -0.97***    & <0.001     & [-1.29, -0.67] \\
{\bf Text In Image}                  & -0.29*      & 0.046      & [-0.62, 0.01] \\
{\bf Invalid HTML}                    &  1.11***    & <0.001     & [0.65, 1.54] \\
{\bf Long Invisible Random Strings}   &  0.94       & 0.107      & [-2.85, 3.29] \\ 
{\bf Invisible Text}                  &  0.85       & 0.144      & [-1.29, 4.68] \\ 
{\bf Excluded Text}                   & -0.15       & 0.384      & [-0.49, 0.21] \\
{\bf Word Splitting with HTML Tags}   &  0.08       & 0.813      & [-0.68, 0.83] \\
{\bf Capitalized HTML Tags}           &  0.01       & 0.983      & [-0.84, 0.78] \\
{\bf Homograph Attack}                & -0.91       & 0.480      & [-2.94, -0.62] \\ 
{\bf Payload in Attachment}           &  0.04       & 0.973      & [-2.47, 0.33] \\
\midrule
\multicolumn{4}{l}{R² = 0.486, F(10, 375) = 35.43, p < 0.001} \\
\multicolumn{4}{l}{Significance codes : *** p < 0.001, ** p < 0.01, * p < 0.05} \\
\bottomrule
\end{tabular}
\end{table}

\section{Discussion}
\label{sec:discussion}

\subsection{Lessons learned}

Our quantitative analysis of email body obfuscation techniques yields several crucial implications for enhancing anti-phishing defenses. The high prevalence observed for diverse methods — ranging from embedding content in images ({\bf Text in Image}) and using standard encodings ({\bf Base64 Encoding}) to deliberately breaking HTML structure ({\bf Invalid HTML}) or hiding content ({\bf Invisible Text}, {\bf Long Invisible Random Strings}) — underscores that effective protection demands robust, multi-modal analysis capabilities operating directly on the email body. Defenses reliant solely on basic text scanning are demonstrably insufficient; they must integrate tolerant yet discerning HTML parsers, reliable decoding mechanisms, statistical content analysis, and effective image processing (e.g., OCR) to comprehensively inspect the varied ways attackers conceal malicious payloads.

Furthermore, the statistically significant co-occurrence patterns revealed, such as the frequent pairing of {\bf Text in Image} with {\bf Excluded Text} (Multipart Abuse), provide compelling evidence that attackers often employ layered, strategic combinations rather than isolated techniques. This highlights the need for detection logic that moves beyond simply flagging individual indicators. Developing rulesets or machine learning models trained to recognize these empirically observed, synergistic combinations and assign appropriately higher risk scores is essential for countering more sophisticated, multi-pronged evasion attempts.

The effectiveness association analysis using SpamAssassin scores, while specific to one filter configuration, offers valuable preliminary insights. The finding that techniques like {\bf Base64 Encoding} and {\bf Text in Image} were associated with score reductions suggests these methods may indeed challenge certain types of filter rules and thus warrant focused detection efforts. Conversely, the fact that {\bf Invalid HTML} significantly increased spam scores indicates that penalizing structural anomalies can be a valid defensive tactic, at least for some filters. Defenses could potentially leverage such correlations to refine risk scoring, contingent on broader validation across diverse filtering platforms. Given the resource constraints inherent in detection systems, prioritizing robust countermeasures against the most prevalent and potentially evasive techniques identified (e.g., {\bf Text in Image}, {\bf Base64 Encoding}, {\bf Invalid HTML}, {\bf Excluded Text}) represents a practical approach to addressing the bulk of observed threat tactics.

Finally, the dynamic nature of phishing necessitates continuous vigilance. The techniques and combinations favored by attackers inevitably evolve. Therefore, regularly performing quantitative analyses of obfuscation trends, similar to the methodology employed here, is vital for adapting defenses, updating detection signatures and heuristics, and retraining machine learning models to maintain their efficacy against the shifting evasion landscape. An empirically grounded understanding of how attackers manipulate email body content and structure is fundamental to building resilient anti-phishing capabilities.

\subsection{Limitations} 
This study, while providing novel quantitative insights, has limitations that should be acknowledged:

\begin{itemize}
    \item \textbf{Dataset Specificity:} The findings are based on 386 emails collected from a specific source (SignalSpam) during a particular period in 2024. The prevalence and patterns of obfuscation may differ in other datasets reflecting different timeframes, geographical origins, or campaign types. Generalizability requires validation on larger, more diverse corpora.
    \item \textbf{Manual Feature Extraction:} The identification and labeling of techniques were performed manually by trained analysts. While ensuring high accuracy for this dataset and enabling nuanced discovery, this process is labor-intensive, limiting scale, and potentially subject to interpretation biases, although cross-validation was used to mitigate this.
    \item \textbf{Effectiveness Analysis Scope:} Our assessment of filter effectiveness relies solely on a single configuration of SpamAssassin (v4.0.1, with a particular ruleset date) and employs association analysis. This implies that the observed results may diverge significantly when compared to other anti-spam systems, especially those that are commercially available and leverage machine learning, or even when different versions or rule sets of SpamAssassin are utilized. Furthermore, it is crucial to understand that the regression analysis reveals statistical relationships, not concrete causal relationships indicative of evasion. A negative coefficient suggests a potential association with reduced scores within this specific filter context, but does not guarantee the efficacy of evasion across diverse systems or situations.
    \item \textbf{Focus on Email Body:} This study deliberately focused on obfuscation within the email body. It did not analyze email headers, sender reputation, URL characteristics (beyond homographs), or landing page features, which are also crucial components of phishing attacks and defenses.
\end{itemize}


\section{Conclusion and future work}
\label{sec:conclusion}
This paper presented a quantitative analysis of body obfuscation techniques in phishing emails, focusing on their prevalence, combination patterns, and association with spam filter scores. Addressing a gap in the empirical understanding of layered evasion tactics, we analyzed 386 manually verified phishing emails to characterize attacker strategies within the email body.

Our key quantitative contributions provide a foundational empirical baseline:
\begin{enumerate}
    \item We established the prevalence of various body obfuscation techniques, confirming the frequent use of methods like {\bf Text in Image} (47.0\%) and {\bf Base64 Encoding} (31.2\%), alongside significant usage of structural manipulation ({\bf Invalid HTML}, 28.8\%; {\bf Excluded Text}, 21.0\%) and content hiding ({\bf Invisible Text}, 15.3\%; {\bf Long Invisible Random Strings}, 15.1\%).
    \item We provided robust statistical evidence (via co-occurrence analysis) for non-random, strategic pairings of techniques, revealing common synergistic combinations such as {\bf Text in Image} with {\bf Excluded Text} (Multipart Abuse) and {\bf Invalid HTML} with various content hiding methods.
    \item We conducted an initial quantitative assessment of technique association with filter scores using SpamAssassin (v4.0.1), identifying techniques significantly associated with score reduction ({\bf Base64 Encoding}, {\bf Text in Image}) and score increase ({\bf Invalid HTML}) within this specific test configuration, offering preliminary indicators of their potential impact on certain filter types.
    \item The diversity and combination of techniques observed demonstrate that attackers often target multiple detection modalities (e.g., text analysis, structural parsing, image analysis) concurrently within a single email.
\end{enumerate}

The primary contribution lies in the quantitative framework employed, moving beyond descriptive cataloging to statistically analyze technique prevalence, co-occurrence, and filter score associations. Our findings offer quantifiable insights into the multifaceted evasion strategies employed by phishers, demonstrating systematic combinations targeting different potential defense layers. This data-driven understanding is vital for the cybersecurity community, highlighting the limitations of defenses focused narrowly on individual indicators and underscoring the need for detection systems capable of recognizing complex, combinatorial patterns. The empirical grounding presented here provides a crucial foundation for future research into characterizing evolving phishing tactics and for engineering more adaptive and resilient anti-phishing solutions. Understanding the attacker's empirically measured strategy is a critical step toward defeating it.

This research opens several promising directions for future work. First, these findings should be validated and extended by utilizing larger, more diverse email datasets collected from varied sources and across different time periods. Second, it is important to develop and evaluate reliable automated methods for extracting body obfuscation features, which would enable large-scale, continuous analysis. Third, future investigations should examine the actual impact of specific obfuscation patterns on deliverability, specifically correlating them with inbox placement rates across major email providers. Fourth, systematic analyses are needed to integrate these body techniques with other factors like email headers, attachment characteristics, URL features, and landing page tactics, thereby providing a holistic view of evasion strategies. Finally, exploring the effectiveness of these techniques against modern, machine learning-based enterprise email security solutions is crucial.


\begin{credits}
\subsubsection{\ackname} 
The authors would like to thank Institut en Cybersécurité D'Occitanie and CNRS for their support in funding this research.

\subsubsection{\discintname} 
The authors declare that they have no competing interests relevant to the content of this article.
\end{credits}

%
%
%
\bibliographystyle{splncs04} 
\bibliography{bib} 

\end{document}